\documentclass[12pt]{article}
\usepackage{graphicx}

\def\leaderfill{\leaders\hbox to 1em{\hss.\hss}\hfill}  %% точка-проводник
\begin{document}

\noindent {\footnotesize\it  ISSN 1063-7737,
 Astronomy Letters, 2008, Vol. 34, No. 10, pp. 686--698.
 \copyright Pleiades Publishing, Inc., 2008.
 \noindent Original Russian Text
 \copyright V.V. Bobylev, 2008, published in Pis'ma v
Astronomicheski$\check{\imath}$ Zhurnal, 2008, Vol. 34, No. 10,
pp. 757--770.}

\noindent
\begin{tabular}{llllllllllllllllllllllllllllllllllllllllllllll}
& & & & & & & & & & & & & & & & & & & & & & & & & & & & & & & & & & & & &  \\
\hline \hline
\end{tabular}

\vskip 1.5cm

 \centerline {\large\bf Open Clusters IC 4665 and Cr 359 and a Probable Birthplace}
 \centerline  {\large\bf of the Pulsar PSR B1929+10}
 \bigskip
 \centerline {V.V. Bobylev}
 \bigskip
 \centerline {\small\it
Central (Pulkovo) Astronomical Observatory of RAS, St-Petersburg}
 \bigskip

{\bf Abstract--}Based on the epicyclic approximation, we have
simulated the motion of the young open star clusters IC 4665 and
Collinder 359. The separation between the cluster centers is shown
to have been minimal 7 Myr ago, 36 pc. We have established a close
evolutionary connection between IC 4665 and the Scorpius-Centaurus
association --- the separation between the centers of these
structures was $\approx200$~pc 15 Myr ago. In addition, the center
of IC 4665 at this time was near two well-known regions of coronal
gas: the Local Bubble and the North Polar Spur. The star HIP 86768
is shown to be one of the candidates for a binary (in the past)
with the pulsar PSR B1929+10. At the model radial velocity of the
pulsar $V_r= 2\pm50$~km s$^{-1}$, a close encounter of this pair
occurs in the vicinity of IC 4665 at a time of $-1.1$~Myr. At the
same time, using currently available data for the pulsar B1929+10
at its model radial velocity $V_r=200\pm50$~km~s$^{-1}$, we show
that the hypothesis of Hoogerwerf et al. (2001) about the breakup
of the $\zeta$Oph---B1929+10 binary in the vicinity of Upper
Scorpius (US) about 0.9 Myr ago is more plausible.

\section*{INTRODUCTION}

At present, fairly reliable estimates of such parameters as the
distance and proper motion components are available for a number
of nearby neutron stars, but as yet there is no information about
their radial velocities. In the problem of determining the
possible birthplace of a neutron star, the radial velocity is a
free parameter.

Based on the Hipparcos Catalogue and radioastrometric data,
Hoogerwerf et al. (2001) analyzed 56 nearby high-velocity stars
($r<700$~pc), the so-called runaway stars, to determine their
possible birthplace inside nearby young stellar groups --- open
star clusters or associations. One of their conclusions was that
the star $\zeta$ Oph and the pulsar PSR B1929+10 (which, according
to the J2000 coordinate system, was designated as PSR J1932+1059)
originated jointly in a supernova explosion in a binary about 1
Myr ago in the part of the Scorpius-Centaurus association
designated as US (Upper Scorpius).

However, subsequent VLBI measurements of the parallax for PSR
B1929+10 (Brisken et al. 2002; Chatterjee et al. 2004) showed that
the probable distance to the pulsar differs significantly from
that used by Hoogerwerf et al. (2001). In the opinion of
Chatterjee et al. (2004), the new data basically reject the
hypothesis of Hoogerwerf et al. (2001) about the birthplace of PSR
B1929+10 at the boundary of the Scorpius-Centaurus association and
its connection with the star $\zeta$ Oph. In this paper, we test
Hoogerwerf's hypothesis using new observational data. In addition,
estimating the probability of other hypotheses is also of
considerable interest.

One of such hypotheses was proposed by Walter (2001), who
suggested considering a different neutron star as a binary (in the
past) with $\zeta$ Oph, namely, RX J185635-3754.

For our part, we also suggest considering the hypothesis that the
birthplace of PSR B1929+10 could be the open cluster IC 4665 or Cr
359. The idea of this hypothesis is based on the simulations of
the PSR B1929+10 trajectory over the celestial sphere (Fig. 2 from
Hoogerwerf et al. (2001)) at various radial velocities (+200 or
$-$200 km s$^{-1}$), which showed that it passed in the immediate
vicinity of the open clusters IC 4665 and Cr 359. In this paper,
we determine the radial velocity of the pulsar at which its
encounter with these clusters is closest.

In the immediate solar neighborhood with a radius of 200--300 pc,
we know such regions of interstellar tenuous hot gas with a
temperature $T\approx10^6$~K as the Local Bubble and the North
Polar Spur. Since the Local Bubble is a compact region that is
almost free from absorbing matter, it was first discovered by
Fitzgerald (1968) through an analysis of the interstellar
reddening of stars. The absorption maps of the distribution of
neutral gas in NaI lines constructed by Sfeir et al. (1999) show
an inclination to the Galactic plane as for the Gould Belt. An
overview of the formation scenarios for the North Polar Spur can
be found in Miller et al. (2007). Data on all of the known bubbles
in the solar neighborhood were collected in Heiles (1998).

The physical processes that produce bubbles, in outline, are as
follows. Multiple supernova explosions occur in young open
clusters. This gives rise to stellar winds and shock waves that
sweep out the gas to the periphery of the local region in
question, where it produces blobs in the form of shells or
extended walls. At the shell boundaries, the gas density increases
significantly, the gas cools down, and molecular clouds are
formed. During supernova explosions, the gas inside the shell can
heat up and radiate in the X-ray range. In the opinion of
Bergh\"ofer and Breitschwerdt (2002), the most realistic theory
for the origin of the Local Bubble is the hypothesis about
multiple, but not simultaneous explosions of about 20 supernovae
during the past 10--20 Myr. Frish (1995), Maiz-Apell\'aniz (2001),
and Breitschwerdt and de Avillez (2006) substantiate the viewpoint
that supernova explosions in the Scorpius-Centaurus association,
which is part of the Gould Belt, are the most suitable objects
responsible for the formation of both the Local Bubble and the
North Polar Spur.

The open clusters IC 4665 and Cr 359 have a high probability of
belonging to the Gould Belt both in their age and spatial position
(Piskunov et al. 2006) and in kinematics (Bobylev 2006). Since
there are very few young open star clusters in the first Galactic
quadrant that are members of the Gould Belt and that have reliable
kinematic data (particularly the radial velocities), studying the
trajectories of the open clusters IC 4665 and Cr 359 and members
of the Scorpius-Centaurus association is of great importance for
understanding the evolution of the Gould Belt.

The goal of this paper is to establish the evolutionary
connections of the young open clusters IC 4665 and Cr 359 with the
structure of the Gould Belt, to establish the possible influence
of the clusters on the bubble development, and to determine the
most probable birthplace of the pulsar PSR B1929+10 by varying its
radial velocity.

The problems posed here are solved by constructing the
trajectories of the objects using the epicyclic approximation.

\section*{THE METHOD}

In this paper, we use a rectangular Galactic coordinate system
with the axes directed away from the observer toward the Galactic
center $(l=0^\circ, b=0^\circ$, the $X$ axis), along the Galactic
rotation $(l=90^\circ, b=0^\circ$, the $Y$ axis), and toward the
North Galactic Pole $(b=90^\circ$, the $Z$ axis). The
corresponding space velocity components of the objects $U,V,W$ are
also directed along the $X,Y,Z$ axes (Kulikovskii 1985).

The method of the epicyclic approximation (Lindblad 1927, 1959)
allows the orbits of stars to be constructed in a coordinate
system rotating around the Galactic center in a circular orbit.We
assume that the origin of the coordinate system coincides with the
local standard of rest and that the stars move along epicycles in
the direction opposite to the Galactic rotation. We use the method
in a form associated with a rectangular coordinate system (Fuchs
et al. 2006):
$$
\displaylines{\hfill
 X(t)= X(0)+{U(0)\over \kappa} \sin(\kappa t)+{V(0)\over 2B} (1-\cos(\kappa t)), \hfill\llap(1)\cr\hfill
 Y(t)= Y(0)+2A \biggl( X(0)+{V(0)\over 2B} \biggr) t
       - {\Omega_0\over B\kappa} V(0) \sin(\kappa t)
       +{2\Omega_0\over \kappa^2} U(0) (1-\cos(\kappa t)),\hfill\cr\hfill
 Z(t)= {W(0)\over \nu} \sin(\nu t) + Z(0) \cos(\nu t), \hfill
 }
$$
where $t$ is the time in Myr (pc/Myr=0.978 km s$^{-1}$), which is
measured into the past; $\kappa=\sqrt{-4\Omega_0 B}$ is the
epicyclic frequency; A and B are the Oort constants; $\Omega_0$ is
the angular velocity of Galactic rotation of the local standard of
rest, $\Omega_0=A-B$; and $\nu=\sqrt{4\pi G \rho_0}$ is the
frequency of the vertical oscillations, where $G$ is the
gravitational constant and $\rho_0$ is the star density in the
solar neighborhood. The space velocities of the objects are
calculated for any necessary time using the formulas
$$ \displaylines{\hfill
 U(t)= U(0) \cos(\kappa t)-{\kappa\over -2B} V(0) \sin(\kappa t), \hfill\llap(2)\cr \hfill
 V(t)= {-2B\over \kappa} U(0) \sin(\kappa t) + V(0) \cos(\kappa t),\hfill\cr \hfill
 W(t)= W(0) \cos(\nu t) - Z(0)\nu \sin(\nu t). \hfill
 }
$$
The parameters $X(0),Y(0),Z(0)$ and $U(0),V(0),W(0)$ in Eqs. (1)
and (2) denote the current positions and velocities of the
objects. The velocities $U,V,W$ are given relative to the local
standard of rest with $(U,V,W)_{LSR}=(10.00,5.25,7.17)\pm(0.36,
0.62,0.38)$ km s$^{-1}$ (Dehnen and Binney 1998).

Following Fuchs et al. (2006), we adopted
 $\rho_0=0.1~M_\odot $pc$^{-3}$,
 which gives
 $\nu=0.074$ km s$^{-1}$ pc$^{-1}$. We also took the Oort constants
 $A= 13.7\pm0.6$ km s$^{-1}$ kpc$^{-1}$ and
 $B=-12.9\pm0.4$ km s$^{-1}$ kpc$^{-1}$ that we
found previously (Bobylev 2004) by analyzing the independent
determinations of these parameters by various authors; in that
case, $\kappa=0.037$ km s$^{-1}$ pc$^{-1}$.

\section*{THE DATA}

The necessary input data on the objects under consideration, such
as their equatorial coordinates, proper motion components, radial
velocities, and parallaxes, are given in Table 1. Table 2 gives
the spatial coordinates
 $X(0),Y(0),Z(0)$ and space velocities
 $U(0),V(0),W(0)$ of the stars and clusters under consideration.

 {\bf IC 4665 and Cr 359}.
  For the open cluster IC 4665, we use the
coordinates of the center, proper motion components, and an
estimate of the photometric distance $r=352\pm70$ pc from the
CRVOCA Catalog (Kharchenko et al. 2007) and the radial velocity
from Manzi et al. (2007), where it was determined using 39 most
probable cluster members. According to Manzi et al. (2007), the
age of IC 4665 is $27.7^{+4.2}_{-3.5}$ Myr. This estimate is
interesting in that it depends little on the quality of
isochrones, because it was obtained by comparing the lithium
abundances in cluster stars with such well-known young clusters as
NGC 2547 and IC 2391. Other authors give the following
``isochronic'' cluster age estimates: 36 Myr (Mermilliod 1981),
30--100 Myr (Prosser 1993), and 43 Myr (Piskunov et al. 2006).

For the open cluster Cr 359, we use the coordinates of the center,
proper motion components, and radial velocity from the CRVOCA
Catalog (Kharchenko et al. 2007) and the distance $r=450\pm200$~pc
from Lodieu et al. (2006), who conducted a critical review of the
estimates obtained by various authors. According to Lodieu et al.
(2006), the age of Cr 359 is $60\pm20$ Myr. Other authors estimate
this cluster as a younger one, 32 Myr (Piskunov et al. 2006) and
$\approx30$~Myr (Wielen 1971; Abt and Cardona 1983).

According to Kharchenko et al. (2007), the apparent radii of IC
4665 and Cr 359 are $1.^\circ0$ and $1.^\circ1$, respectively.

de Wit et al. (2006) estimated the mass of IC 4665 to be 300--350
$M_\odot$. Cr 359 probably has a similar mass.

{\bf PSR B1929+10}. The radio pulsar PSR B1929+10 is an isolated
neutron star with an age of 3 Myr (estimated from the ratio P/2\.P
and belongs to the population of nearby neutron stars that are
closely related to the Gould Belt (Popov et al. 2003; Motch et al.
2006).

For PSR B1929+10, we use the data from Chatterjee et al. (2004)
that were obtained from VLBI measurements. The most important
refined parameter is the new parallax of the pulsar, $\pi=2.77
\pm0.07$~mas.

Note that Hoogerwerf et al. (2001) adopted a model value of
$\pi=4\pm2$~mas. The preliminary distance estimates for the pulsar
that made it closer to the Sun served as a basis for this choice.

{\bf US, UCL, and LCC}. The initial positions of the centers
 $(X(0),Y (0),Z(0))=(134,-20,52)$ pc and heliocentric
velocities
 $(U(0),V (0),W(0))=(-6.7,-16.0,-8.0)\pm(5.9, 3.5,2.7)$ km s$^{-1}$ for US,
 $(119,-67,31)$~pc and
 $(-6.8,-19.3,-5.7)\pm(4.6,4.7,2.5)$ km s$^{-1}$ for UCL,
 $(62,-100,10)$~pc and
 $(-8.2,-18.6,-6.4)\pm(5.1,7.3,2.6)$~km s$^{-1}$ for LCC
 were taken
from Fern\'andez et al. (2006). These authors obtained the
positions of the centers (with errors of 1.2 pc) using data from
de Zeeuw et al. (1999); the velocities were determined by Sartori
et al. (2003) based on the space velocities of a large number of
stars (more than 120 for each of the groups). Fern\'andez et al.
(2006) used these data to analyze the kinematics of the members of
the Scorpius. Centaurus association. However, the orbits of the
objects were calculated using a model that, apart from the
axisymmetric Galactic potential, also included the potential from
a spiral density wave and the potential of the bar at the Galactic
center (Fern\'andez et al. 2008). Comparing the results obtained
by various methods is of considerable interest.

\section*{RESULTS}
 \subsection*{The Neighborhoods of IC 4665, Cr 359, and PSR B1929+10}

{\bf The encounters of the centers}.
 Based on Eqs. (1) and (2), we
obtained several solutions at various radial velocities of the
pulsar PSR B1929+10, $V_r$ (PSR). The results are the following:

(1) at $V_r$ (PSR) = $+45$ km s$^{-1}$, the pulsar encounters with
the cluster IC 4665 at the distance
 $\Delta_r=\sqrt{\Delta X^2+\Delta Y^2+\Delta Z^2}=52$~pc at
 $t=-0.85$~Myr;

(2) at $V_r$ (PSR) = $-60$ km s$^{-1}$, the minimum distance from
the pulsar to the center of Cr 359 is $\Delta_r=38$~pc at
$t=-1.0$~Myr;

(3) at $V_r$ (PSR) = 0 km s$^{-1}$, we obtain a joint encounter
with the two clusters. In this case, the minimum distances from
the pulsar to the centers of Cr 359 and IC 4665 are
$\Delta_r=69$~pc and $\Delta_r=60$~pc, respectively, at
$t=-1.0$~Myr.

As we see, the pulsar trajectory can pass through the coronas of
the two clusters under consideration. As was noted by de Wit et
al. (2006), the tidal radius of IC 4665 is $\approx1^\circ$. At an
assumed distance to the cluster of $0r=352$~pc, it is
$\approx6$~pc. However, the stars with a common proper motion
occupy an area of $\approx100$ square degrees on the celestial
sphere (Lodieu et al. 2006) and, hence, the coronal radius of IC
4665 is $\approx30$~pc. As can be seen from Table 2, the error in
the spatial position of the center of IC 4665 is about $\pm50$~pc;
its spatial displacement in 2 Myr is insignificant.

In this respect, Cr 359 is of lesser interest, since the error in
its distance is 44\%. Therefore, the radius of its corona lies
within the range 20--60 pc, but the spatial localization accuracy
of the cluster is low, about $\approx150$ pc (Table 2).

Next, we take the next natural step or, more specifically, we
search for a suitable rapidly flying star, a possible member of a
binary. For this purpose, we use the list of such stars from
Hoogerwerf et al. (2001). We can see from Fig. 2 in Hoogerwerf et
al. (2001) that the trajectories of three stars passed in the
immediate vicinity of IC 4665 and Cr 359 on the celestial sphere
with a radius of $\approx15^\circ$ 2 Myr ago: HIP 66524, HIP
86768, and HIP 91599. To calculate the space velocities, we invoke
the current radial velocities of these stars from the Pulkovo
Catalog of Radial Velocities (Gontcharov 2006).

The construction of the trajectories for the three rapidly flying
stars shows that HIP 86768 is most interesting among them.

Indeed, HIP 66524 moves away fairly rapidly toward the Galactic
center in $X$ coordinate, without having a close encounter with
the objects of interest to us.

For HIP 91599 (star No. 20 from the list of Hoogerwerf et al.
2001), the situation is slightly different. At the radial velocity
of the pulsar PSR B1929+10 $V_r$ (PSR)=$+250$ km s$^{-1}$, it
encounters with HIP 91599 at the distance $\Delta_r=47$~pc at
$t=-0.5$ Myr. The modulus of the velocity difference between the
pulsar and HIP 91599 at the time of their encounter is
$\Delta_V=\sqrt{\Delta U^2+\Delta V^2+\Delta W^2}=316$~km
s$^{-1}$.

At the radial velocity of the pulsar $V_r$ (PSR)=$+2$ km s$^{-1}$,
it encounters with HIP 86768 at the distance $\Delta_r=19$~pc at
$t=-1.1$ Myr. In this case, the modulus of the velocity difference
between the pulsar and HIP 86768 at the time of their encounter
$\Delta_V=206$ km s$^{-1}$. Taking into account the fairly large
errors in the distances and velocities of HIP 86768, we can assume
that a closer encounter is also possible. Therefore, simulating
the encounters by taking into account the errors in the data is of
great interest.

{\bf Monte Carlo Simulations of Encounters}.
 (1) Initially, we
directly repeated the numerical experiment of Hoogerwerf et al.
(2001) on the encounter between the star $\zeta$ Oph and PSR
B1929+10. We computed 3 million orbits by taking into account the
random errors in the input data, which were normally distributed
with the $3\sigma$ region. For the pulsar, we used old data:
 $V_r (PSR) =200\pm50$ km s$^{-1}$,
 $\pi=4\pm2$ mas,
 $\mu_\alpha\cos\delta= 99\pm12~(6\times2)$ mas yr$^{-1}$, and
 $\mu_\delta= 39\pm8~(4\times2)$~mas yr$^{-1}$ (following Hoogerwerf et al. (2001),
 we doubled the initial errors
in the proper motions, as shown in parentheses). The results are:
out of the 3 million orbits, 32505 encounters occur at distances
$\Delta_r\leq10$~pc (1.1\%); in 4410 of the 32505 cases, the star
and the pulsar were no farther than 10 pc from the US center about
1 Myr ago. The derived parameters are in excellent agreement with
the results of Hoogerwerf et al. (2001).

(2) The same experiment on the encounter between $\zeta$ Oph and
PSR B1929+10 with the currently available data for the pulsar:
 $V_r$ (PSR) = $200\pm50$ km s$^{-1}$,
 $\pi=2.77\pm0.7(0.07\times10)$~mas,
 $\mu_\alpha\cos\delta= 94.09\pm3.3 (0.11\times30)$~mas yr$^{-1}$, and
 $\mu_\delta= 42.99\pm4.8 (0.16\times30)$~mas yr$^{-1}$.
 The initial
values of the random errors (as shown in parentheses) were
increased in order to have conditions comparable to those in
experiment (1). The results are: out of the 3 million orbits,
74115 encounters occur at distances $\Delta_r\leq10$~pc (2.5\%);
in 5611 of the 74115 cases, the star and the pulsar were no father
than 10 pc from the US center about 1 Myr ago. The domains of
admissible values of $V_r$, $\pi$,
 $\mu_\alpha\cos\delta~(\mu^*_\alpha)$, and
 $\mu_\delta$ at which 5611 encounters occur
are presented in Fig. 1 for the star $\zeta$ Oph and in Fig. 2 for
PSR B1929+10. Figure 3 shows the expected distribution F3D of
minimum distance $\Delta_r$ calculated from the formula of
Hoogerwerf et al. (2001),
$$ \displaylines{\hfill
 F_{3D}(\Delta_r)=
 {\Delta^2_r\over {2\sigma^3\sqrt\pi}}\exp\biggl[-{\Delta^2_r\over{4\sigma^2}}\biggr] \hfill\llap(3)
 }
$$
for the adopted $\sigma=2.5$~pc.

(3) The experiment on the encounter between HIP 86768 and PSR
B1929+10 for $V_r$ (PSR) = $2\pm50$ km s$^{-1}$ and the same
values of
 $\pi= 2.77\pm0.7 (0.07\times10)$ mas,
 $\mu_\alpha\cos\delta= 94.09\pm3.3 (0.11\times30)$ mas yr$^{-1}$, and
 $\mu_\delta= 42.99\pm4.8 (0.16\times30)$ mas yr$^{-1}$
 as those in experiment~(2).
 The results are: out of the 3 million orbits, 22708 encounters occur at distances $\Delta_r\leq10$~pc
(0.8\%); in 6932 of the 22708 cases, the star and the pulsar were
no father than 80 pc from the center of IC 4665 about 1 Myr ago.
The domains of admissible values of $V_r$, $\pi$,
 $\mu_\alpha\cos\delta~(\mu^*_\alpha)$, and
 $\mu_\delta$ at which 6932 encounters occur are presented in Fig. 4
for HIP 86768 and in Fig. 5 for PSR B1929+10. Figure 6 shows the
expected distribution of minimum distance $\Delta_r$ calculated
from Eq.~(3) for the adopted $\sigma=6$~pc.

(4) The experiment of the simulation of encounters between HIP
91599 and PSR B1929+10 at $V_r$ (PSR) = $250\pm50$ km s$^{-1}$
shows that out of the 3 million orbits, 32961 encounters occur in
the range 15 pc$<\Delta_r\leq30$~pc (0 encounters at
$\Delta_r\leq15$~pc) about 0.5 Myr ago. The stars HIP 86768 and
HIP 91599 do not encounter at distances $\Delta_r\leq10$~pc,
suggesting that the triple system HIP 86768---HIP 91599---B1929+10
is unlikely.

(5) Testing the hypothesis of Walter (2001) --- the experiment on
the encounter between $\zeta$ Oph and the isolated neutron star
(NS) RX J185635-3754 at $V_r$ (NS)= $-50\pm50$ km s$^{-1}$. The
results are: out of the 3 million orbits, 2144 encounters occur at
distances $\Delta_r\leq10$~pc (0.07\%); there are no encounters
with the US center at $\Delta_r\leq10$~pc. The derived
characteristics are in good agreement with the results of
Hoogerwerf et al. (2001).

\subsection*{IC 4665 and the Scorpius-Centaurus Association}

We traced the trajectories of IC 4665 and Cr 359 in a time
interval comparable to their lifetime, up to $t=-30$ Myr. For the
mean values of the input parameters described in Section 2.1, we
found the minimum separation between the cluster centers to be
$\Delta_r=36$ pc at $t=-7$ Myr. Allowance for the errors in the
cluster distances and velocities showed that the encounter time
lies within the interval 0--12 Myr.

Figure 7 shows the positions of IC 4665, Cr 359, and members of
the Scorpius-Centaurus association and their trajectories in the
past 30 Myr; since the age of US does not exceed 5 Myr, the
corresponding part of the trajectory is marked by the dotted line.

Figure 8 shows the calculated positions of IC 4665, the members of
the Scorpius-Centaurus association UCL and LCC, and two bubbles,
the Local Bubble (LB) and the North Polar Spur (L1) (Breitschwerdt
and de Avillez 2006) at $t=-15$ Myr. The difference in the
positions of the UCL and LCC centers at this time found from
comparison with the data of Ortega et al. (2002) or Fern\'andez et
al. (2006) is $\approx\pm50$ pc along the $X$ and $Y$ axes and
very small, $\approx\pm3$ pc, along the $Z$ axis. We think these
discrepancies attributable to differences in the models to be
insignificant for the purposes of this paper. Note the approach of
Fuchs et al. (2006), who simulated not the trajectories of the
group centers, but the individual trajectories of very massive
stars, potential supernovae, using Eq. (1). Some of them turned
out to be near bubble L1 at $t=-15$~Myr (see Fig. 8). For IC 4665,
apart from the coordinates of the center, Fig. 8 shows a cloud of
300 points (distributed with the $3\sigma$ region) calculated by
taking into account the errors in the input data for the cluster.
As can be seen from the figure, the contribution from the errors
in the distance is dominant.

\section*{DISCUSSION}
\subsection*{The Connection of IC 4665 and Cr 359 with the
Scorpius-Centaurus Complex}

As we see from Fig. 7, the edge of the Gould Belt located in the
first Galactic quadrant continuously approached the Galactic plane
in a time interval of $\approx30$ Myr in the past. This is in good
agreement with the model for the evolution of the Gould Belt
suggested by Olano (2001) and is consistent with the model
calculations of the kinematic evolution of the Gould Belt
performed by Perrot and Grenier (2003) using data on molecular
clouds for the scenario with a ``Galactic-plane crossing'' (Fig. 7
from Perrot and Grenier (2003)), whereby the age of the Gould Belt
is $51.8\pm1.0$ Myr. According to these two models, the edge of
the Gould Belt located in the first Galactic quadrant reaches the
Galactic plane at $\approx-10$ Myr. Our results show a similar
motion. Thus, for example, the $Z$ coordinates of the centers of
UCL and LCC (which lie in the plane of the Gould Belt) at
$t=-15$~Myr fixed in Fig. 7 are close to 0 pc, while the $Z$
coordinates of the centers of IC 4665 and Cr 359 (rising above the
plane of the Gould Belt) are close to 70 pc.

Most of the authors believe that the Scorpius-Centaurus
association is part of the Gould Belt (de Zeeuw et al. 1999;
Bobylev 2006). However, Fern\'andez et al. (2008) advocate the
point of view that the association could evolve independently of
the Gould Belt (outside the scope of the supernova explosion
hypothesis) and could be formed from the parent cloud that was
compressed by collisions with a spiral density wave.

On the other hand, whatever the origin of the association, the
history of star formation in it agrees satisfactorily with the
model of successive star formation suggested by Blaauw (1964;
1991) and developed by Preibish and Zinnecker (1999), as applied
to US. The age estimates for the members of the Scorpius-Centaurus
association (de Geus et al. 1989) lie within the ranges 5--6 Myr
for US, 14--15 Myr for UCL, and 11--2 Myr for LCC. Both the
present-day ``isochronic'' estimates (Mamajek et al. 2002; Sartori
et al. 2003), 8--10 Myr for US and 16--20 Myr for UCL and LCC, and
our kinematic age estimate for the entire association, 21 Myr
(Bobylev and Bajkova 2007), agree well with these estimates.

Our results show that the separation between IC 4665 and the
Scorpius-Centaurus association was much smaller in the past. Thus,
for example, whereas the current separation between their centers
is 302 pc (UCL--IC 4665), it was 215 pc 15 Myr ago and 120 pc 30
Myr ago. The minimum separation was $\approx80$ pc at $-66$ Myr,
but our model approximation on such a long time scale is already
unreliable. All of these facts suggest that they were formed from
the same parent hydrogen cloud; the cluster IC 4665 has always
been located on its periphery.

According to the model of successive star formation, the shocks
from supernova explosions compress the nearby (within $\approx100$
pc) molecular clouds, triggering star formation in them. In our
view, the role of IC 4665 in the evolution of the
Scorpius-Centaurus association could lie in the fact that
supernova explosions in this cluster could trigger star formation
processes in the association itself.

At $t=-15$ Myr, IC 4665 was near two well-known regions of coronal
gas: the Local Bubble and the North Polar Spur.

Based on an analysis of HI data, Heiles (1998) obtained the
following parameters for the North Polar Spur (which is known as
the brightest component of radio loop I): the heliocentric
distance to the center is 120 pc, the coordinates of the center
are $l=320^\circ$ and $b=+5^\circ$, and the radius is 118 pc.
According to the model of Willingale et al. (2003), it can be
represented as an expanding spherical superbubble with the
following parameters: the distance is 210 pc, the direction of the
center is $l=352^\circ$, $b=+15^\circ$, and the bubble radius is
140 pc. X-ray observations of the North Polar Spur showed that the
peak of emission with an energy of 3/4 keV is observed in the
direction with $l=26.^\circ8$ and $b=+22.^\circ0$ (Snowden et al.
1997; Willingale et al. 2003; Miller et al. 2007), i.e., at the
bubble boundary. The current coordinates of IC 4665 are
 $l=30.^\circ6$ and $b=+17.^\circ1$. Taking into account the results shown in Figs. 7
and 8, we conclude that IC~4665 has always been located at a
distance of 150--200 pc from the bubble boundary both in the past
and at present. This means that supernova explosions in IC 4665
over the past several Myr could influence the formation of the
North Polar Spur. This influence lies in the fact that supernova
explosions in IC 4665 could produce a stellar counter-wind (the
main explosions in the Scorpius-Centaurus association), which
additionally compresses the bubble walls.

\subsection*{The Probable Birthplace of the Pulsar PSR B1929+10}

We considered the hypothesis of Hoogerwerf et al. (2001) that the
star $\zeta$ Oph and the pulsar B1929+10 could be components of a
binary in the neighborhood (within less than 10 pc) of US about 1
Myr ago using currently available data for the pulsar. Our
simulations of encounters showed that, contrary to the view of
Chatterjee et al. (2004), improved data for the pulsar only
strengthen this hypothesis. Therefore, it seems most plausible.
This requires that the pulsar radial velocity fall within the
range $V_r=200\pm50$ km s$^{-1}$.

Our hypothesis of the HIP 86768---B1929+10 binary, which could
exist about 1 Myr ago in a wide neighborhood (within less than 80
pc) of IC4665, also seems plausible. The probability of the
existence of such a binary is lower than that in the hypothesis of
Hoogerwerf et al. (2001). A distinctly different pulsar radial
velocity, $V_r=2\pm50$ km s$^{-1}$, is required for our hypothesis
to hold.

An examination of the hypothesis by Walter (2001) about the
$\zeta$ Oph---RXJ185635-3754 binary showed its probability to be
an order of magnitude lower than that of our hypothesis.

\section*{CONCLUSIONS}

Based on the epicyclic approximation, we simulated the motion of
the young open star clusters IC 4665 and Collinder 359. The
separation between the cluster centers was found to have been
minimal 7 Myr ago, 36 pc. This suggests that IC 4665 and Collinder
359 were formed from the same parent hydrogen cloud.

We showed a close evolutionary connection between IC 4665 and the
Scorpius-Centaurus association. Thus, for example, the separation
between the centers of these structures was $\approx200$ pc about
15 Myr ago. At that time, IC 4665 was near two well-known regions
of coronal gas: the Local Bubble and the North Polar Spur. This
means that supernova explosions in IC 4665 over the past 15 Myr
could affect the development of bubbles and, in particular, the
North Polar Spur.

Analysis of the parameters for the encounter of the pulsar PSR
B1929+10 with various stars in the vicinity of IC 4665 and Cr 359
suggests that the star HIP 86768 is a suitable candidate for a
binary (in the past) with the pulsar. Monte Carlo simulations of
the encounters of this pair with the pulsar radial velocity
 $V_r=2\pm50$ km s$^{-1}$ showed that out of the 3 million orbits, 22708
encounters occur at distances $\Delta_r<10$ pc (0.8\%); in 6932 of
the 22708 cases, the star and the pulsar were no farther than 80
pc from the center of IC 4665 about 1 Myr ago. We showed that
using currently available data for the pulsar increases the
probability of the hypothesis by Hoogerwerf et al. (2001) about
the breakup of the $\zeta$ Oph---B1929+10 binary in the immediate
vicinity of US about 1 Myr ago. Monte Carlo simulations of the
encounters of this pair with the pulsar radial velocity
$V_r=200\pm50$ km s$^{-1}$ showed that out of the 3 million
orbits, 74115 encounters occur at distances $\Delta_r<10$~pc
(2.5\%); in 5611 of the 74115 cases, the star and the pulsar were
no farther than 10~pc from the US center about 1 Myr ago.

\section*{ACKNOWLEDGMENTS}

I am grateful to A.T. Bajkova for help in the work and to the
referees for several useful remarks that contributed to an
improvement of the paper. This work was supported by the Russian
Foundation for Basic Research
 (project nos. 05-02-17047 and 08-02-00400).

\newpage
\section*{REFERENCES}

{\small

~~~~~1. H. A. Abt and O. Cardona, Astrophys. J. 272, 182 (1983).

2. T.W. Bergh\"ofer and D. Breitschwerdt, Astron. Astrophys. 390,
299 (2002).

3. A. Blaauw, Ann. Rev. Astron. Astrophys. 2, 213 (1964).

4. A. Blaauw, The Physics of star Formation and Early Stellar
Evolution, Ed. by C. J. Lada and N. D. Kylafis (Kluwer, Dordrecht,
1991).

5. V. V. Bobylev, Pis'ma Astron. Zh. 30, 185 (2004) [Astron. Lett.
30, 159 (2004)].

6. V. V. Bobylev, Pis'ma Astron. Zh. 32, 906 (2006) [Astron. Lett.
32, 816 (2006)].

7. V. V. Bobylev and A. T. Bajkova, Pis'ma Astron. Zh. 33, 643
(2007) [Astron. Lett. 33, 571 (2007)].

8. D. Breitschwerdt and M. A. de Avillez, Astron. Astrophys. 452,
L1 (2006).

9. W. F. Brisken, J. M. Benson, W. M. Goss, et al., Astrophys. J.
571, 906 (2002).

10. S. Chatterjee, J. M. Cordes, W. H. T. Vlemmings, et al.,
Astrophys. J. 604, 339 (2004).

11. W. Dehnen and J. J. Binney, Mon. Not. R. Astron. Soc. 298, 387
(1998).

12. D. Fern\'andez, F. Figueras, and J. Torra, Highlights of
Spanish Astrophysics IV Ed. by F. Figueras et al.
(Springer-Verlag, New York, 2006); astro-ph: 0611766v1 (2006).

13. D. Fern\'andez, F. Figueras, and J. Torra, astro-ph:
0801.0605v1 (2008).

14 . M. P. Fitzgerald, Astron. J. 73, 983 (1968).

15. P. C. Frish, Space Sci. Rev. 72, 499 (1995).

16. B. Fuchs, D. Breitschwerdt,M. A. Avilez, et al., Mon. Not. R.
Astron. Soc. 373, 993 (2006).

17. E. J. de Geus, P. T. de Zeeuw, and J. Lub, Astron. Astrophys.
216, 44 (1989).

18. G. A. Gontcharov, Pis'ma Astron. Zh. 32, 844 (2006) [Astron.
Lett. 32, 759 (2006)].

19. C. Heiles, Astrophys. J. 498, 689 (1998).

20. R. Hoogerwerf, J. H. J. de Bruijne, and P. T. de Zeeuw,
Astron. Astrophys. 365, 49 (2001).

21. N. V. Kharchenko, R.-D. Scholz, A. E. Piskunov, et al.,
Astron. Nachr. 328 (2007).

22. P. G. Kulikovskii, Zvezdnaya astronomiya (Stellar Astronomy)
(Nauka, Moscow, 1985) [in Russian].

23. B. Lindblad, Ark. Mat., Astron. Fys. 20 A, No. 17 (1927).

24. B. Lindblad, Handbuch der Physik 53, 21 (1959).

25. N. Lodieu, J. Bouvier, D. J. James, et al., Astron. Astrophys.
450, 147 (2006).

26. J. Maiz-Apell\'aniz, Astrophys. J. 560, L83 (2001).

27. E. E. Mamajek, M. Meyer, and J. Liebert, Astron. J. 124, 1670
(2002).

28. S. Manzi, S. Randich, W. J. de Wit, et al., astro-ph/
0712.0226v1 (2007).

29. J. C. Mermilliod, Astron. Astrophys. 97, 235 (1981).

30. E. D. Miller, H. Tsunemi, M. W. Bautz, et al., astro-ph/
0708.4227v1 (2007).

31. C. Motch, A. M. Pires, F. Haberl, et al., Astrophys. Space
Sci. 308, 217 (2006).

32. C. A. Olano, Astron. Astrophys. 121, 295 (2001).

33. V. G. Ortega, R. de la Reza, E. Jilinski, et al., Astrophys.
J. 575, L75 (2002).

34 .C. A. Perrot and I. A. Grenier, Astron. Astrophys. 404, 519
(2003).

35. A. E. Piskunov, N. V. Kharchenko, S. R\"{o}ser, et al.,
Astron. Astrophys. 445, 545 (2006).

36. S. B. Popov,M. Colpi,M. E. Prokhorov, et al., Astron.
Astrophys. 406, 111 (2003).

37. T. Preibish and H. Zinnecker, Astron. J. 117, 2381 (1999).

38. C. F. Prosser, Astron. J. 105, 1441 (1993).

39. M. J. Sartori, J. R. D. L\'epine, and W. S. Dias, Astron.
Astrophys. 404, 913 (2003).

40. D. M. Sfeir, R. Lallement, F. Grifo, et al., Astron.
Astrophys. 346, 785 (1999).

41. S. L. Snowden, R. Egger, M. J. Freyberg, et al., Astrophys. J.
485, 125 (1997).

42. The Hipparcos and Tycho Catalogues, ESA SP-1200, (1997).

43. F.M.Walter, Astrophys. J. 549, 433 (2001).

44. R.Wielen, Astron. Astrophys. 13, 309 (1971).

45. R. Willingale, A. D. P. Hands, R. S. Warwick, et al., Mon.
Not. R. Astron. Soc. 343, 995 (2003).

46. W. J. de Wit, J. Bouvier, F. Palla, et al., Astron. Astrophys.
448, 189 (2006).

47. P. T. de Zeeuw, R. Hoogerwerf, J. H. J. de Bruijne, et al.,
Astron. J. 117, 354 (1999).

}

\newpage
%%%%%%%%%%%%%%%%%%%%%%%%%%%%%%%%%%%%%%%%%%%%%%
{
\begin{table}[t]
\caption[]{\small\baselineskip=1.0ex\protect
 Data on the objects

}
\begin{center}
\begin{tabular}{|c|c|c|c|c|c|c|c|c|c|c|}\hline
           &&&&&\\
    Object & $\alpha_{(J2000.0)},$ &  $\mu_\alpha\cos \delta,$ & $\mu_\delta,$ & $\pi,$ & $V_r,$ \\
           & $\delta_{(J2000.0)}$  &    mas yr$^{-1}$  &  mas yr$^{-1}$  &  mas & km s$^{-1}$  \\\hline
  B1929+10 & $19^h 32^m 13^s.9497$       & $94.09\pm0.11$ & $ 42.99\pm0.16$ & $2.77\pm0.07$ & --- \\
           & $10^\circ 59' 32''.4203$     &&&&  \\\hline
   IC~4665 & $17^h 46^m$                  & $-0.57\pm0.30$ & $ -7.40\pm0.36$ & $2.84\pm0.56$ & $-16.0\pm1.1$ \\
           & $ 5^\circ 43'$               &&&&  \\\hline
    Cr~359 & $18^h 01^m$                  & $ 0.22\pm0.28$ & $ -8.90\pm0.26$ & $2.22\pm0.99$ & $ -4.6\pm0.2$ \\
           & $ 2^\circ 54'$               &&&& \\\hline
 HIP~66524 & $13^h 38^m 09^s.0095$        & $-60.75\pm0.49$ & $-13.34\pm0.67$ & $2.28\pm0.75$ & $ 23\pm1$ \\
           & $-50^\circ 20' 59''.702$     &&&& \\\hline
 HIP~81377     & $16^h 37^m 09^s.5378$    & $13.07\pm0.85$ & $25.44\pm0.72$ & $7.12\pm0.71$ & $  -9.9\pm5.5$ \\
 ($\zeta$ Oph) & $-10^\circ 34' 01''.524$ &&&&\\\hline
 HIP~86768 & $17^h 43^m 47^s.0205$        & $-7.12\pm0.70$ & $-10.39\pm0.51$ & $2.34\pm0.80$ & $ 19.0\pm4.3$ \\
           & $-7^\circ 04' 46''.588$      &&&& \\\hline
 HIP~91599 & $18^h 40^m 48^s.0517$        & $-9.64\pm1.13$ & $-22.64\pm0.79$ & $3.61\pm1.16$ & $  29 \pm5$   \\
           & $-8^\circ 43' 07''.688$      &&&&  \\\hline
 {\tiny RXJ185635-3754}
           & $18^h 56^m 35^s.56$          & $326.7\pm0.8$ & $-59.1\pm0.7$ & $16.5\pm2.3$ & --- \\
           & $-37^\circ 54' 37''.0$       &&&&  \\\hline
\end{tabular}
\end{center}
\end{table}
}

\newpage
%%%%%%%%%%%%%%%%%%%%%%%%%%%%%%%%%%%%%%%%%%%%%%
{\begin{table}[t]                                                %% t-2
\caption[]{\small\baselineskip=1.0ex\protect
 Positions and heliocentric velocities of the objects

}
\begin{center}
\begin{tabular}{|c|c|c|c|c|c|c|c|c|c|c|c|}\hline
           &&&&&&\\
    Объект & $X(0),$ & $Y(0),$ & $Z(0),$  & $U(0),$ & $V(0),$ & $W(0),$ \\
           &     pc &     pc &    pc   &  km s$^{-1}$  &  km s$^{-1}$  &  km s$^{-1}$  \\\hline
  B1929+10 & $244\pm6$   & $265\pm7 $  & $-25\pm1 $ & $-109\pm34$ & $ 91\pm37$  & $-107\pm6$ \\
   IC~4665 & $290\pm57$  & $171\pm34$  & $104\pm21$ & $   -6\pm2$ & $-17\pm3$   & $  -9\pm2$ \\
    Cr~359 & $381\pm169$ & $217\pm96$  & $ 99\pm44$ & $    6\pm8$ & $-16\pm14$  & $ -10\pm8$ \\
 HIP~66524 & $279\pm92$ & $-326\pm107$ & $ 90\pm30$ & $-83\pm67$  & $-102\pm57$ & $   1\pm3$ \\
 HIP~81377 & $128\pm13$  & $  14\pm1$  & $ 56\pm6$  & $-11\pm5$   & $  18\pm2$  & $  -1\pm2$ \\
 HIP~86768 & $397\pm136$ & $134\pm46$  & $ 86\pm29$ & $   25\pm7$ & $-18\pm17$  & $   6\pm2$ \\
 HIP~91599 & $253\pm81$  & $113\pm36$  & $ -8\pm3$  & $   39\pm7$ & $-18\pm13$  & $  -3\pm1$ \\
 {\tiny RXJ185635-3754}  & $ 58\pm8$ & $-1.4\pm0.2$ & $-18\pm3$   & $-75\pm48$  & $  20\pm1$ & $-75\pm15$ \\\hline
\end{tabular}
\end{center}
 {\small
Note. The velocities of the pulsar B1929+10 were calculated for
its radial velocity $V_r=2\pm50$ km s$^{-1}$; the velocities of
the pulsar RX J185635.3754were calculated for $V_r=-50\pm50$ km
s$^{-1}$.
 }
\end{table}
}

\newpage
%%%%%%%%%%%%%%%%%%%%%%%%%%%%%%%%%%%%%%%% FIG.1:
\begin{figure}[t]
{
\begin{center}
  \includegraphics[width=145mm]{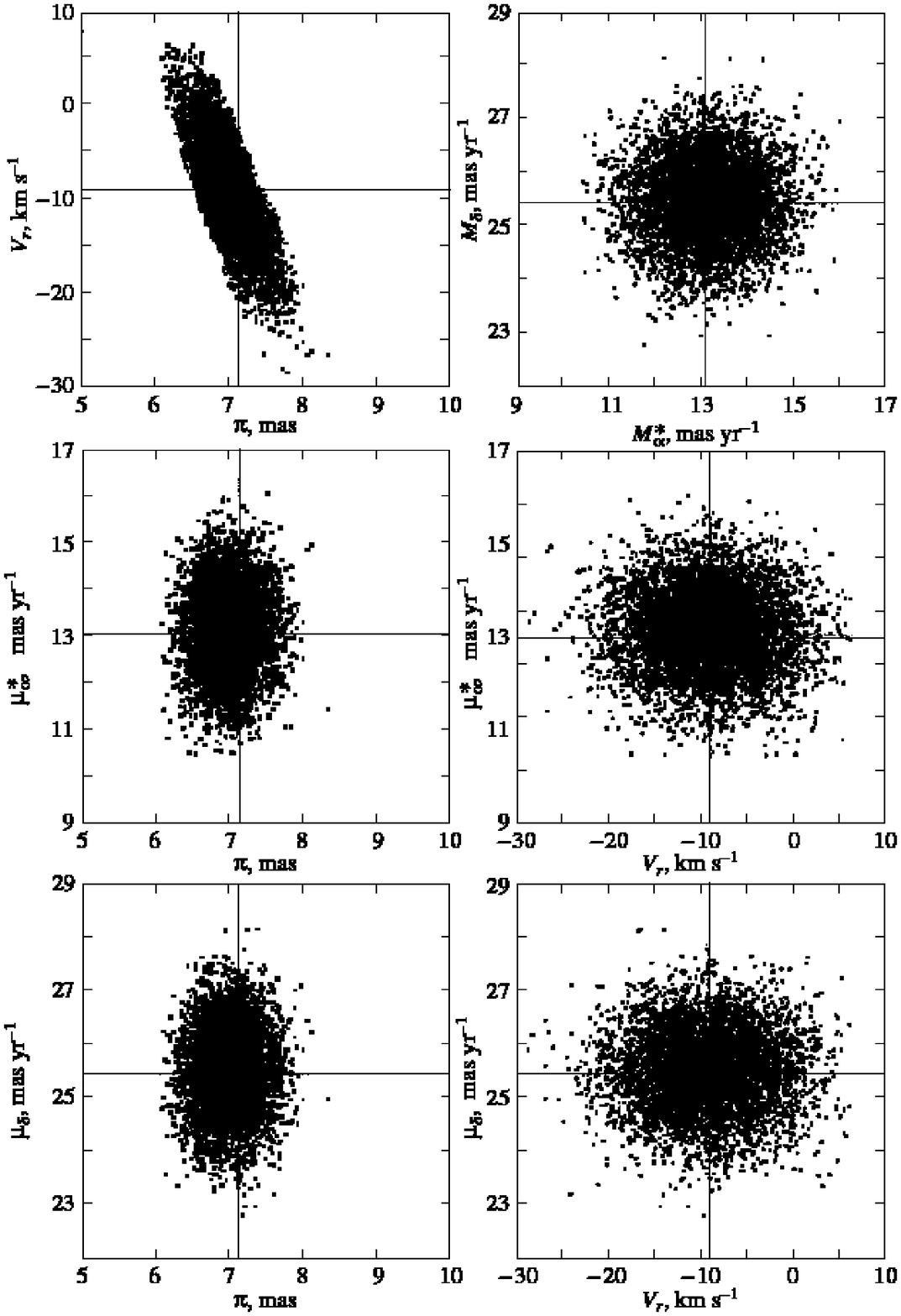}
\end{center}
} Fig.~1. Domains of admissible values at which 5611 encounters
occur at distances $\Delta_r<10$~pc between the star $\zeta$ Oph
and the pulsar PSR B1929+10, during which they fall into the
neighborhood of US, $\Delta_r<10$~pc for $\zeta$ Oph.
\end{figure}
%%%%%%%%%%%%%%%%%%%%%%%%%%%%%%%%%%%%%%%%%%%%%%%%%%%%%%%%%%%%%%%

\newpage
%%%%%%%%%%%%%%%%%%%%%%%%%%%%%%%%%%%%%%%% FIG.2:
\begin{figure}[t]
{
\begin{center}
  \includegraphics[width=140mm]{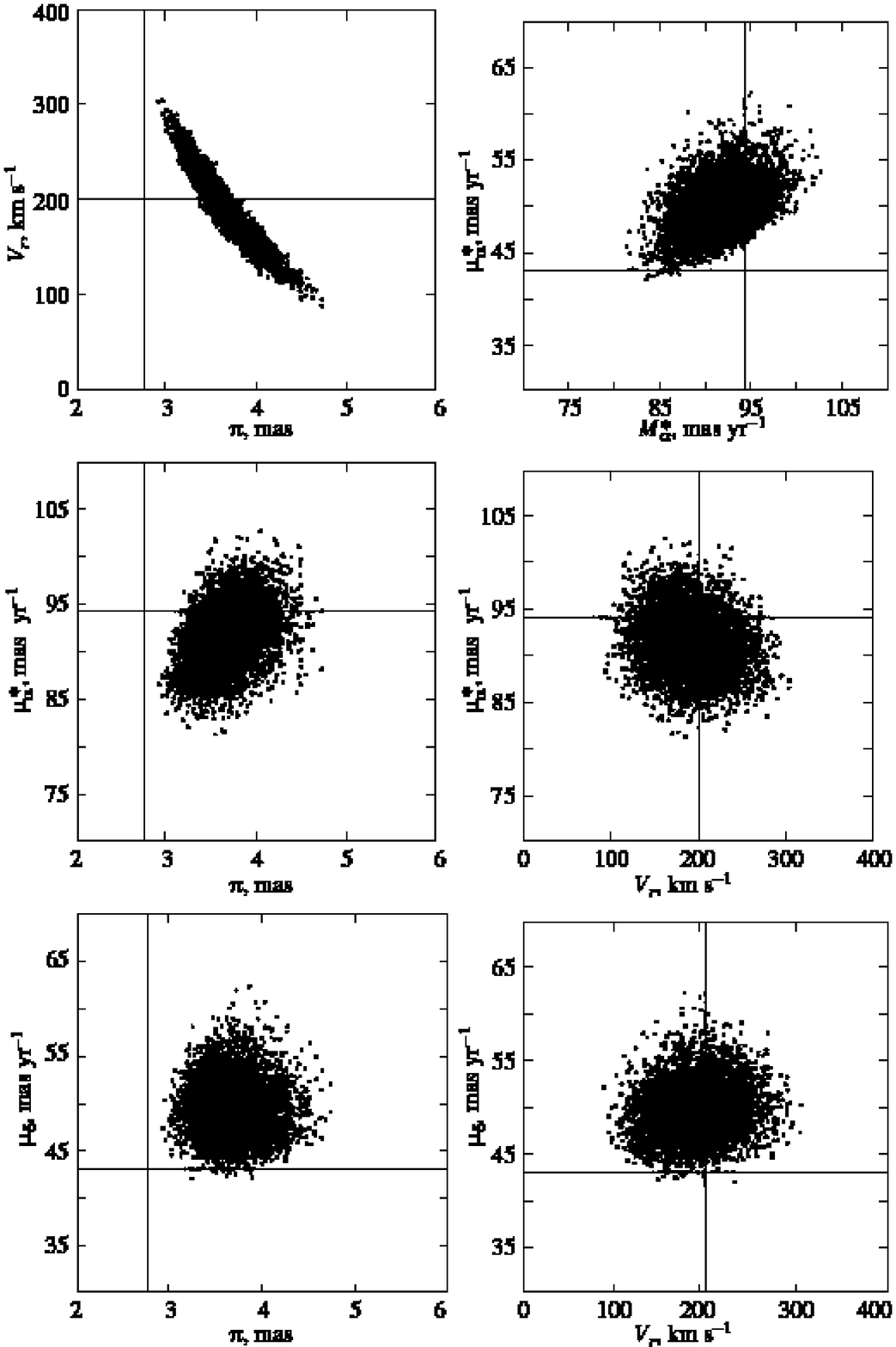}
\end{center}
} Fig.~2. Domains of admissible values at which 5611 encounters
occur at distances $\Delta_r<10$~pc between the star $\zeta$ Oph
and the pulsar PSR B1929+10, during which they fall into the
neighborhood of US, $\Delta_r<10$~pc for PSR B1929+10.
\end{figure}
%%%%%%%%%%%%%%%%%%%%%%%%%%%%%%%%%%%%%%%%%%%%%%%%%%%%%%%%%%%%%%%

\newpage
%%%%%%%%%%%%%%%%%%%%%%%%%%%%%%%%%%%%%%%% FIG.3:
\begin{figure}[t]
{
\begin{center}
  \includegraphics[width=120mm]{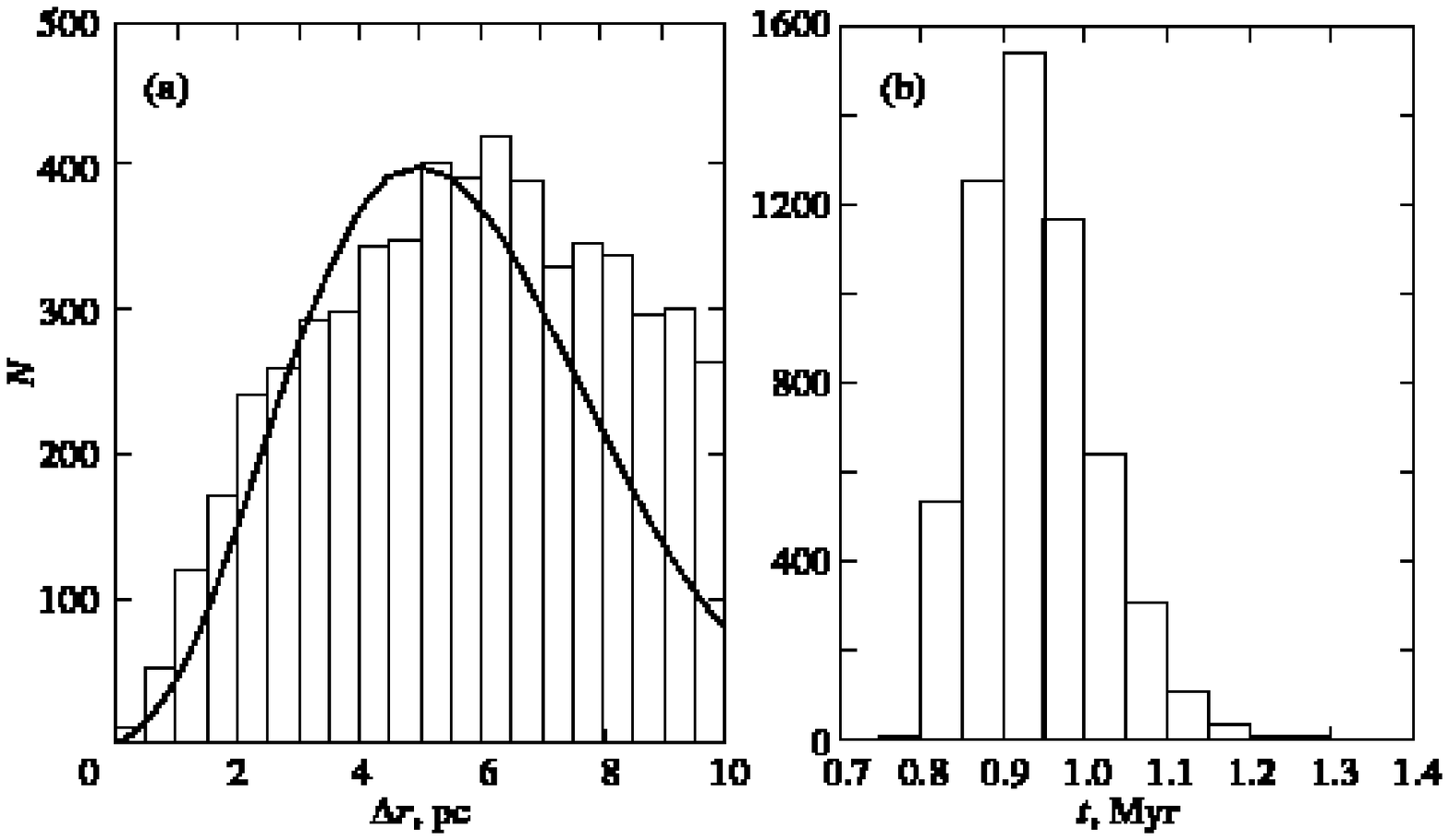}
\end{center}
} Fig.~3. (a) Expected distribution of minimum distance
$\Delta_r<10$~pc for 5611 encounters of $\zeta$~Oph with PSR
B1929+10 and (b) histogram of encounter times.
\end{figure}
%%%%%%%%%%%%%%%%%%%%%%%%%%%%%%%%%%%%%%%%%%%%%%%%%%%%%%%%%%%%%%%

\newpage
%%%%%%%%%%%%%%%%%%%%%%%%%%%%%%%%%%%%%%%% FIG.4:
\begin{figure}[t]
{
\begin{center}
  \includegraphics[width=145mm]{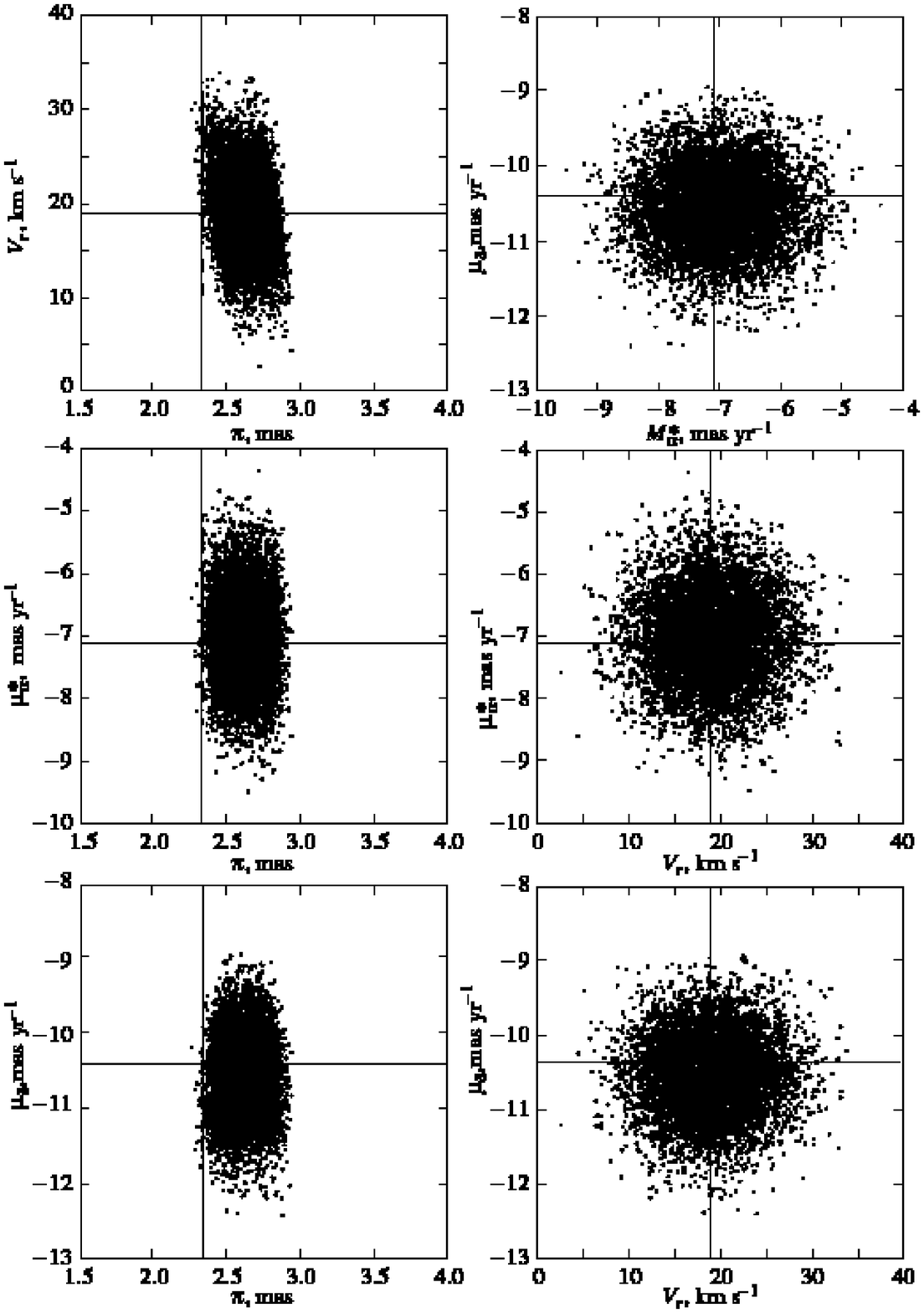}
\end{center}
} Fig.~4. Domain of admissible values at which 6932 encounters
occur at distances $\Delta_r<10$~pc between HIP 86768 and PSR
B1929+10, during which they fall into the neighborhood of IC 4665,
$\Delta_r<80$~pc for HIP 86768.
\end{figure}
%%%%%%%%%%%%%%%%%%%%%%%%%%%%%%%%%%%%%%%%%%%%%%%%%%%%%%%%%%%%%%%

\newpage
%%%%%%%%%%%%%%%%%%%%%%%%%%%%%%%%%%%%%%%% FIG.5:
\begin{figure}[t]
{
\begin{center}
  \includegraphics[width=145mm]{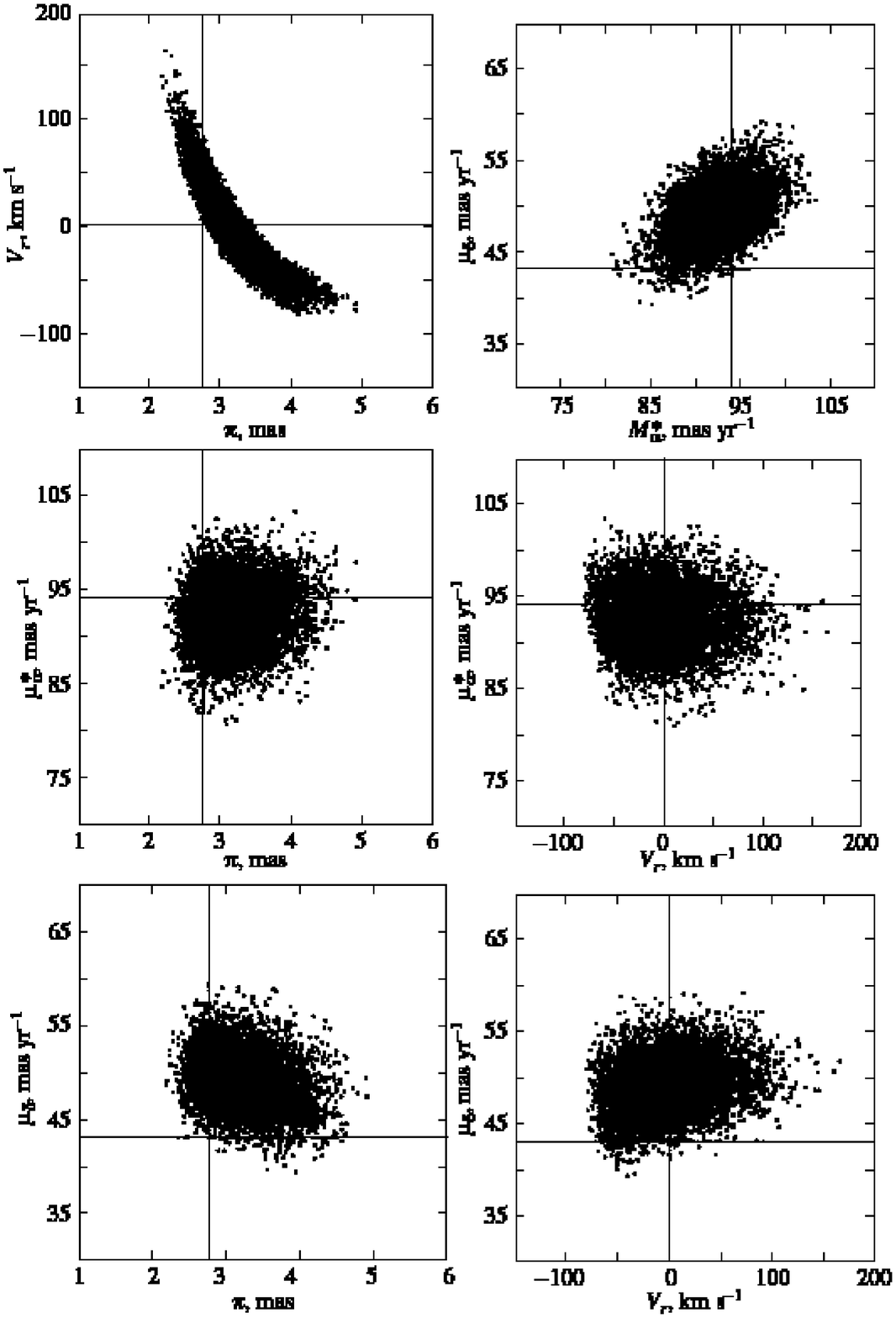}
\end{center}
} Fig.~5. Domains of admissible values at which 6932 encounters
occur at distances $\Delta_r<10$~pc between HIP 86768 and PSR
B1929+10, during which they fall into the neighborhood of IC 4665,
$\Delta_r<80$~pc for PSR B1929+10.
\end{figure}
%%%%%%%%%%%%%%%%%%%%%%%%%%%%%%%%%%%%%%%%%%%%%%%%%%%%%%%%%%%%%%%

\newpage
%%%%%%%%%%%%%%%%%%%%%%%%%%%%%%%%%%%%%%%% FIG.6:
\begin{figure}[t]
{
\begin{center}
  \includegraphics[width=120mm]{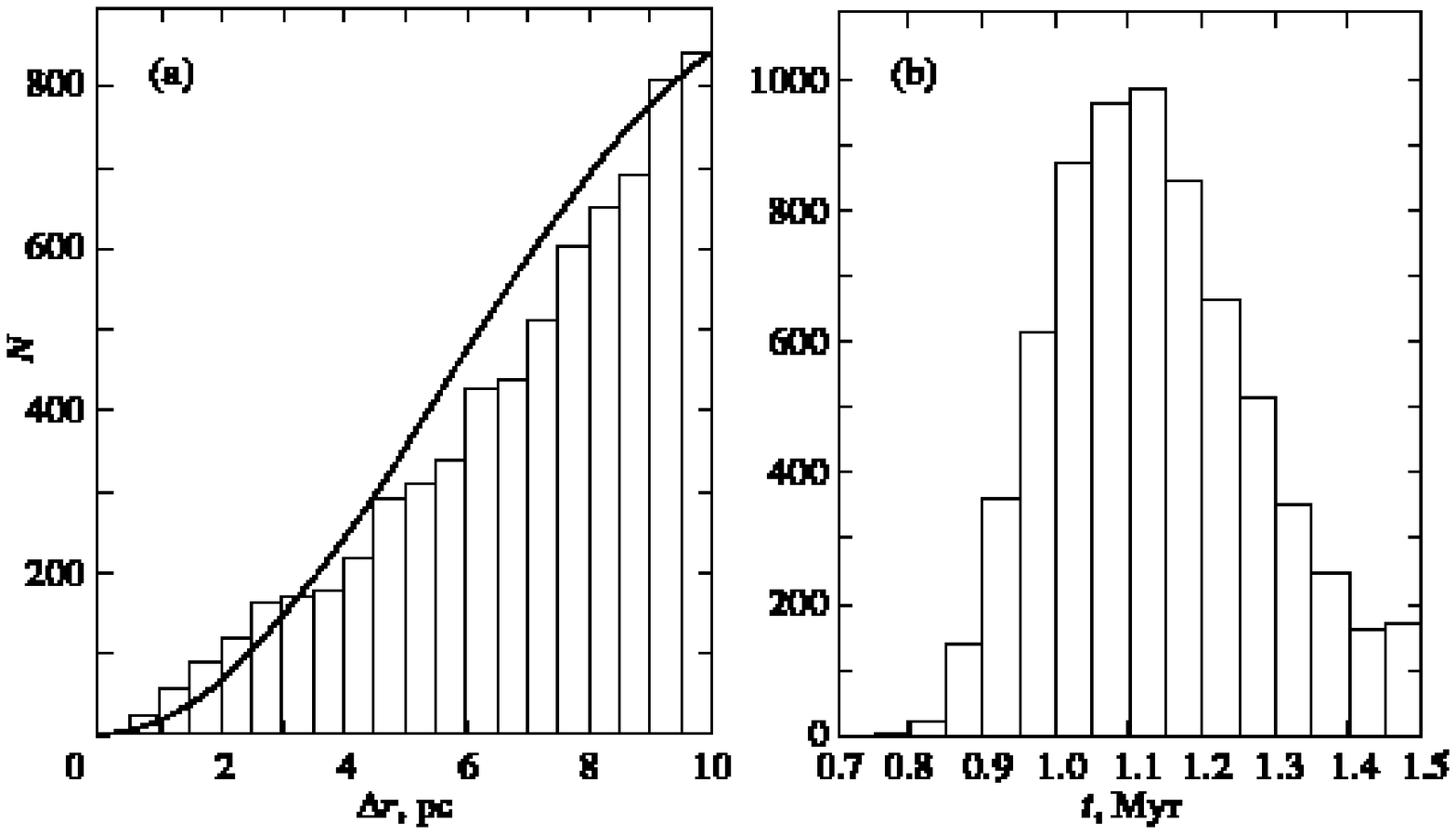}
\end{center}
} Fig.~6. (a) Expected distribution of minimum distance
$\Delta_r<10$~pc for 6932 encounters of HIP 86768 with PSR
B1929+10 and (b) histogram of encounter times.
  \vskip 140mm
\end{figure}
%%%%%%%%%%%%%%%%%%%%%%%%%%%%%%%%%%%%%%%%%%%%%%%%%%%%%%%%%%%%%%%

\newpage
%%%%%%%%%%%%%%%%%%%%%%%%%%%%%%%%%%%%%%%% FIG.7:
\begin{figure}[t]
{
\begin{center}
  \includegraphics[width=90mm]{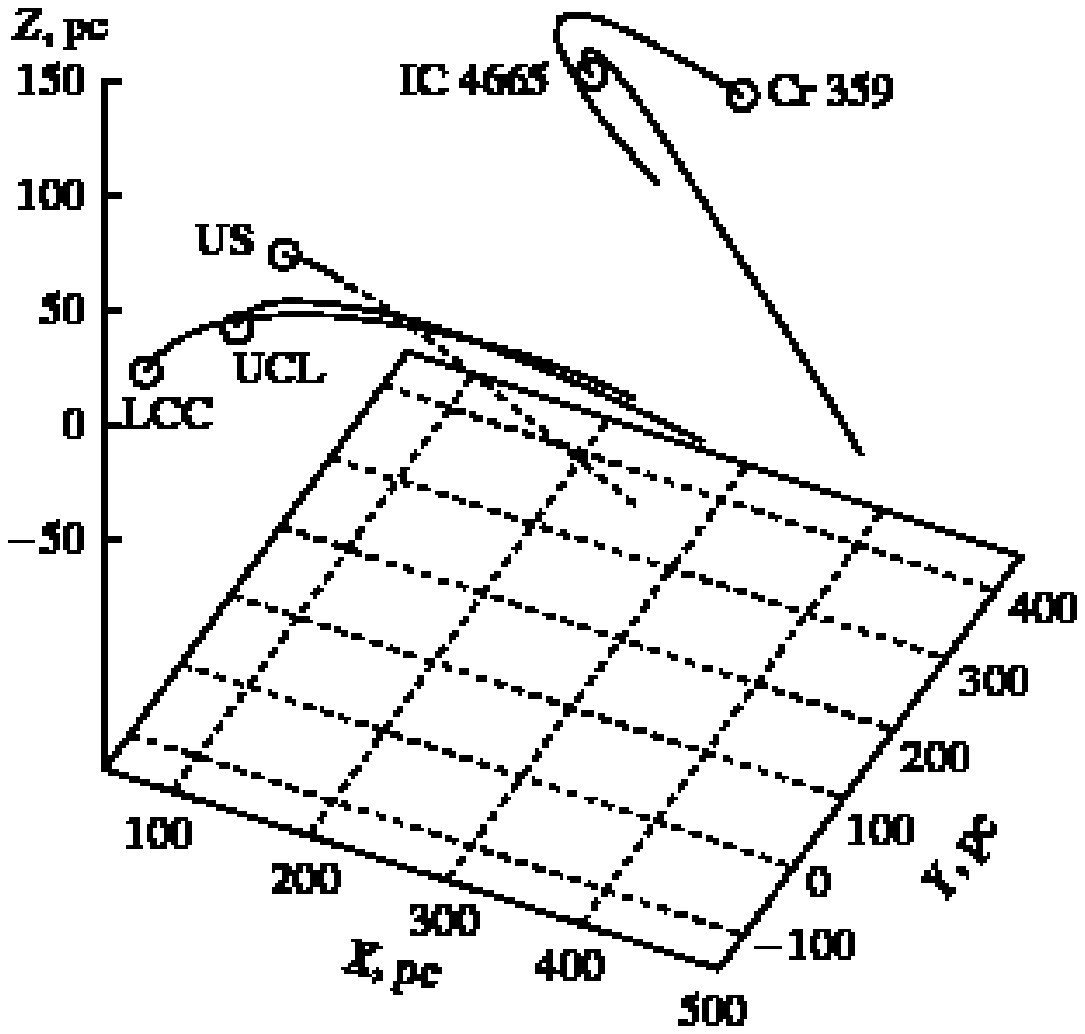}
\end{center}
} Fig.~7. Spatial coordinates of IC 4665, Cr 359, members of the
Scorpius-Centaurus association, and their trajectories in the past
30 Myr.
\end{figure}
%%%%%%%%%%%%%%%%%%%%%%%%%%%%%%%%%%%%%%%%%%%%%%%%%%%%%%%%%%%%%%%

\newpage
%%%%%%%%%%%%%%%%%%%%%%%%%%%%%%%%%%%%%%%% FIG.8:
\begin{figure}[t]
{
\begin{center}
  \includegraphics[width=90mm]{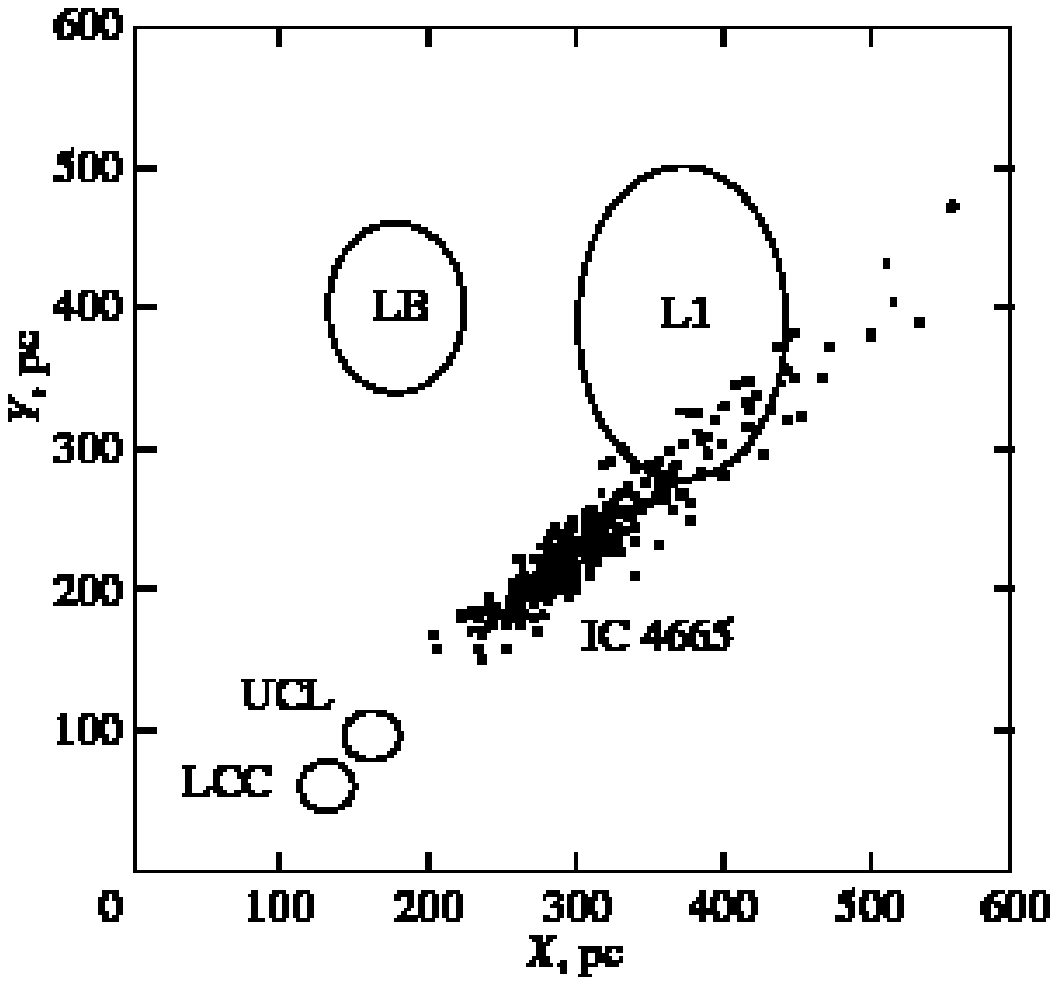}
\end{center}
} Fig.~8. Positions of IC 4665, UCL, LCC, the Local Bubble (LB),
and the North Polar Spur (L1) (Breitschwerdt and de Avillez 2006)
15 Myr ago..
\end{figure}
%%%%%%%%%%%%%%%%%%%%%%%%%%%%%%%%%%%%%%%%%%%%%%%%%%%%%%%%%%%%%%%

\end{document}